\documentclass[aps,pra,twocolumn]{revtex4-1}

\usepackage{amsfonts,amsmath,amssymb,graphicx}

\newcommand{\ket}[1]{| #1\rangle}
\newcommand{\dc}{\mathrm{dc}}
\newcommand{\ac}{\mathrm{ac}}
\newcommand{\ext}{\mathrm{ext}}
\newcommand{\FDC}{F^{(\mathrm{dc})}}
\newcommand{\FAC}{F^{(\mathrm{ac})}}
\newcommand{\unit}[1]{\,\mathrm{#1}}
\DeclareMathOperator{\BJ}{J}
\DeclareMathOperator{\tr}{tr}
\DeclareMathOperator{\CZ}{CZ}

\begin{document}

\title{Flux control of superconducting qubits at dynamical sweet spots}

\author{Nicolas Didier}

\affiliation{Rigetti Computing, 775 Heinz Avenue, 94710 Berkeley, California}

\begin{abstract}
Scaling up superconducting quantum processors with optimized performance requires a sufficient flexibility in the choice of operating points for single and two qubit gates to maximize their fidelity and cope with imperfections. 
Flux control is an efficient technique to manipulate the parameters of tunable qubits, in particular to activate entangling gates. 
At flux sensitive points of operation, the ubiquitous presence of $1/f$ flux noise however gives rise to dephasing by inducing fluctuations of the qubit frequency.
We show how two-tone modulation of the flux bias, a bichromatic modulation, gives rise to a continuum of dynamical sweet spots where dephasing due to slow flux noise is suppressed to first order for a wide range of time-averaged qubit frequencies.
The qubits can be operated at these dynamical sweet spots to realize protected entangling gates and to avoid collisions with two-level-system defects.
\end{abstract}

\maketitle

\section{Introduction}

Manipulating quantum systems while maintaining or improving their coherence is essential to realize high-fidelity quantum gates and cope with experimental imperfections.
This has been one of the main focuses of the field of quantum information processing in superconducting circuits over the last two decades, with the experimental realization of qubits based on 
the Cooper-pair box~\cite{Nakamura:1999,Vion:2002,Wallraff:2004,Koch:2007,Paik:2011,Barends:2014,Arute:2019}, 
using persistent currents~\cite{Mooij:1999,Yan:2016},
built with superinductances~\cite{Manucharyan:2009,Earnest:2018,Nguyen:2019,Pechenezhskiy:2019,Gyenis:2019}, 
and encoding information into continuous variables~\cite{Ofek:2016,Lescanne:2019,Grimm:2019,CampagneIbarcq:2019}.
Flux control is a particularly useful approach for quantum state manipulation
as it allows one to vary the transition frequencies of superconducting qubits.
In the presence of frequency collisions with two-level-system defects~\cite{Klimov:2018,Burnett:2019,Schlor:2019,Lisenfeld:2019,Bilmes:2019} and other qubits,
flux control is used for instance to park the qubits away from such resonances.
To realize entangling gates between two qubits, flux control is used to adjust their detuning or their coupling or both.
Fast DC flux pulses are used to bring qubits into resonance during the gate time~\cite{Strauch:2003,DiCarlo:2010,Reed:2012,Rol:2019,Barends:2019,Arute:2019}.
Another method is to modulate the flux bias of tunable qubits
in order to generate sidebands separated by the modulation frequency around the time-averaged qubit frequency~\cite{Bertet:2006,Niskanen:2007,Beaudoin:2012,Strand:2013,McKay:2016,Naik:2017,Roth:2017,Mundada:2019}.
One sideband is then brought into resonance with a second qubit to generate entanglement~\cite{Matt:2018, Shane:2018, Sab:2019, Deanna:2019b}.
The speed of the gate is determined by the weight of the sideband used in the coherent exchange~\cite{Nico:2018}.

Using SQUID loops instead of single Josephson junctions to build tunable qubits however opens the system to flux noise that constitutes an additional source of dephasing~\cite{Koch:1983,Wellstood:1987,Vion:2002,Martinis:2003,Ithier:2005,
	Yoshihara:2006,Bialczak:2007,KochR:2007,Faoro:2008,Manucharyan:2009,
	Barends:2013,Omalley:2015,Kumar:2016,Yan:2016,Quintana:2017,
	Kou:2017,Plourde:2017,You:2019}.
The structure of flux noise in superconducting circuits is usually composed of a pink noise characterized by a $1/f$ spectral density that dominates at low frequencies, on top of a white noise background.
While flux noise generated by the control electronics can be mitigated with improved apparatus, especially concerning the white noise contribution, 
flux noise emerging from the material is still challenging to suppress.
The coherence-limited infidelity of quantum logical gates in presence of flux noise typically scales as the gate time over the dephasing time.
Fast two-qubit gates are thus required when operated at flux-sensitive points.
When the fidelity is coherence limited, it can still be increased by protecting the qubit from flux noise.
Operating points at which the qubit is weakly sensitive to flux noise, aka sweet spots,
have been experimentally explored under DC flux biases~\cite{Vion:2002} and RF flux modulation~\cite{Sab:2019, Schuyler:2019} for qubits based on the split Cooper pair box.
Reducing the dephasing rate is also possible with bipolar flux pulses~\cite{Rol:2019}.
Under flux modulation, the tunable qubit is sensitive to two kinds of flux noise: the additive flux noise on the DC bias and the multiplicative flux noise on the AC amplitude.
Dynamical sweet spots are achieved at operating points insensitive to both additive and multiplicative flux noise.
Experimentally, dephasing time is usually limited by $1/f$ flux noise.
In presence of slow flux noise the dephasing rate is proportional to the slope of the time-averaged frequency under modulation versus the applied flux,
dynamical sweet spots are thus found at critical points of the time-averaged frequency in the parking-flux--modulation-amplitude plane.
Dynamical sweet spots have also been explored experimentally in electromechanical systems~\cite{Pirkkalainen:2013} and quantum dots~\cite{Frees:2019}.

Sweet spots are however sparse.
For asymmetric transmons, on a flux quantum period $\Phi_0$ the first sweet spots are located at biases $\Phi_\dc=0,\frac{1}{2}\Phi_0$ without modulation and at an amplitude of $\Phi_\ac\approx0.6\,\Phi_0$ with modulation, as well as at a bias $\Phi_\dc\approx\pm\frac{1}{4}\Phi_0$ and an amplitude of $\Phi_\ac\approx0.4\,\Phi_0$~\cite{Nico:2019}.
The five corresponding qubit frequencies (or time-averaged frequency under modulation) are stretched almost evenly over the tunability range.
The implementation of entangling gates protected from flux noise is then tied to few points of operation.
In this work, we show how to obtain a continuum of dynamical sweet spots with a bichromatic modulation, i.e.~by modulating the flux bias with two tones, at a fundamental frequency $f_m$ and one harmonic $pf_m$.
The corresponding effective transmon frequency spans a large part of the tunability range;
it can be optimized, together with the sideband weights, for high-fidelity single and two-qubit gates.
We show how to protect entangling gates on a wide range of operating points.
We explicitly focus on $1/f$ flux noise, even though experimentally the coherence time is limited by the $T_1$ contribution to $T_2$ and other sources of noise.
The abundance of dynamical sweet spots under bichromatic modulation allows for a precious flexibility in the manipulation of protected quantum states, crucial to scale up noisy intermediate-scale quantum processors.

The paper is organized as follows. 
In Sec.~\ref{Sec_sweetspot} we present how to find dynamical sweet spots under bichromatic flux modulation and the corresponding time-averaged frequencies.
In Sec.~\ref{Sec_gates} we focus on entangling gates at dynamical sweet spots obtained by sideband engineering.
In Sec.~\ref{Sec_TLS} we show how to manipulate tunable qubits at dynamical sweet spots, in particular to decouple from two-level-system defects.

\section{Dynamical sweet spots under bichromatic flux modulation}
\label{Sec_sweetspot}

The sensitivity to flux noise under modulation directly results from the properties of the frequency band versus flux bias.
The frequency $f$ of a tunable superconducting qubit is an even function of the flux bias $\Phi_\ext$ that is flux-quantum periodic.
It can be described as a Fourier series $\FDC_{n}$ that depends on the electrical parameters of the qubit~\cite{Nico:2018},
\begin{align}
f(\phi_\ext) &= \sum_{n=0}^\infty\FDC_{n} \cos(n\phi_\ext),
\label{FBA}
\end{align}
where we note the phase bias $\phi_\ext=2\pi\Phi_\ext/\Phi_0$.
For tunable qubits made of an asymmetric SQUID, the coefficients $\FDC_{n}$ decrease rapidly with the index~$n$.
We now consider a bichromatic modulation, characterized by the parking flux $\Phi_\dc$, the modulation amplitude $\Phi_\ac$, the modulation frequency $f_m$ and second tone frequency $pf_m$ ($p\geq2$ integer), the mixing angle $\alpha$ and the phases $\theta_1$, $\theta_p$ of the two tones,
\begin{align}
\Phi_\ext(t) = \; & \Phi_\dc + \Phi_\ac M(t), \label{flux}\\
M(t) = \; & \cos(\alpha)\cos(2\pi f_mt+\theta_1) \nonumber \\
+ \; & \sin(\alpha)\cos(2\pi pf_mt+\theta_p).
\label{signal}
\end{align}
Under this flux modulation, the time evolution of the qubit frequency is conveniently expressed in terms of a Fourier series,
\begin{align}
f(t) &= \bar{f} + \sum_{k=1}^\infty \FAC_{k} \cos(2\pi k f_mt+\theta_k),
\label{fourier}
\end{align}
where the coefficients $\FAC_{k}$ and phases $\theta_k$ are provided in Appendix~\ref{App_fourier}.

\begin{figure}[t]
	\centering
	\includegraphics[width=0.9\columnwidth]{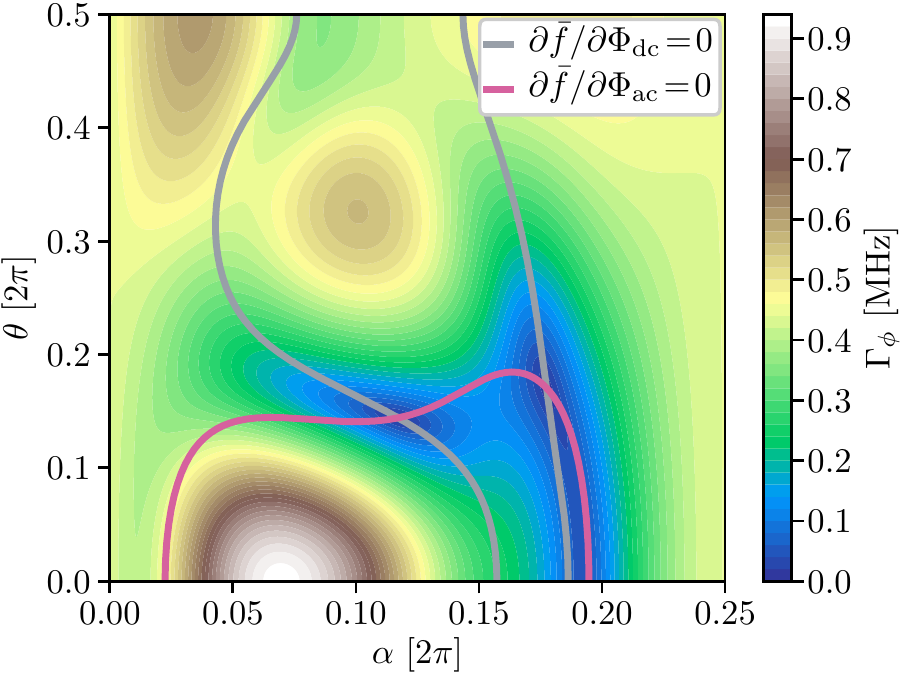}
	\caption{Dephasing rate dependence on the mixing angle $\alpha$ and relative phase $\theta$ of the bichromatic flux modulation  for $p=2$ at $\Phi_\dc=0.125\,\Phi_0$ and $\Phi_\ac=0.75\,\Phi_0$.
		The dynamical sweet spots are found at the intersections between the DC sweet spots where $\partial\bar{f}/\partial\Phi_\dc=0$ (grey lines) and the AC sweet spots where $\partial\bar{f}/\partial\Phi_\ac=0$ (pink line).
		The tunable transmon is characterized by a maximum frequency $f_\mathrm{max}=5\unit{GHz}$, a minimum frequency $f_\mathrm{min}=4.2\unit{GHz}$ and an anharmonicity $\eta_\mathrm{max}=200\unit{MHz}$~\cite{FourierDC}.
		The noise strengths are equal to $A_\dc=A_\ac =33\,\mu\Phi_0$.}
	\label{fig_findsweetspot}
\end{figure}

Under flux modulation the qubit probes the noise spectrum $S(\omega)$ at harmonics of the modulation frequency, $S(k2\pi f_m)$, and the sensitivity is proportional to the slope of the corresponding Fourier coefficient $\FAC_k$ versus flux~\cite{Nico:2019}.
For $1/f$ flux noise the spectral density is usually negligible at RF modulation frequencies and the dephasing rate is proportional, at leading order, to the slope of the time-averaged frequency with respect to the parking flux for additive noise and with respect to the modulation amplitude for multiplicative noise. 
At first order, the total dephasing rate is then equal to
\begin{align}
\Gamma_\phi &= 2\pi\lambda\sqrt{A_\dc^2\left(\frac{\partial\bar{f}}{\partial\Phi_\dc}\right)^2 + A_\ac^2\left(\frac{\partial\bar{f}}{\partial\Phi_\ac}\right)^2},
\label{Gammaphi}
\end{align}
with $A_\dc$, $A_\ac$ the additive and multiplicative $1/f$ noise strength, respectively.
The constant $\lambda=\sqrt{3/2-\gamma-\log(2\pi T_\phi/T_\mathrm{ir})}$, that is typically $\lambda\approx3$,
depends on the infra-red cutoff time $T_\mathrm{ir}$, which is of the order of the measurement time, and $\gamma$ is Euler's constant~\cite{Nico:2019}.
In this derivation we assume that the signal $M(t)$ in Eq.~\eqref{flux} is ideally synthesized such that the global amplification $\Phi_\ac$ and offset $\Phi_\dc$ subsequently applied to generate $\Phi_\ext(t)$ are subjected to flux fluctuations.
The multiplicative flux fluctuations on the  modulation tones thus have the same source, with a strength weighted by the mixing angle~$\alpha$.
The expression of the time-averaged frequency under bichromatic modulation~is
\begin{align}
\bar{f} &= \sum_{m=0}^\infty \cos(m\theta) \sum_{n=1}^\infty \FDC_{n} \cos[n\phi_\dc+(p+1)m\tfrac{\pi}{2}] \nonumber\\ 
&\quad \times (2-\delta_{n,0}) \BJ_{pm}(n\phi_\ac\cos\alpha)\BJ_{m}(n\phi_\ac\sin\alpha),
\label{favg}
\end{align}
where $\theta = \theta_p-p\theta_1$.
Using Chebyshev polynomials, the time-averaged frequency can be written as a polynomial $\bar{f} = P(x)$ in $x\equiv\cos\theta$ ($x\in[-1,1]$).

The dynamical sweet spots are found by looking for the sets of pulse parameters $(\Phi_\dc,\Phi_\ac,\alpha,\theta)$ for which the dephasing rate, Eq.~\eqref{Gammaphi}, vanishes.
Using the symmetries of the time-averaged frequency, it is sufficient to consider $\alpha\in[0,\pi/2]$ and $\theta\in[0,\pi]$.
For a given set of $\Phi_\dc$ and $\Phi_\ac$, the angle $\alpha$ is swept between $0$ and $\pi/2$ to find the intersection between the roots of the polynomials $\partial P(x) / \partial \phi_\dc$ and $\partial P(x) / \partial \phi_\ac$ on the real interval $[-1,1]$.
An example is plotted in Fig.~\ref{fig_findsweetspot}.

\begin{figure}
	\includegraphics[width=\columnwidth]{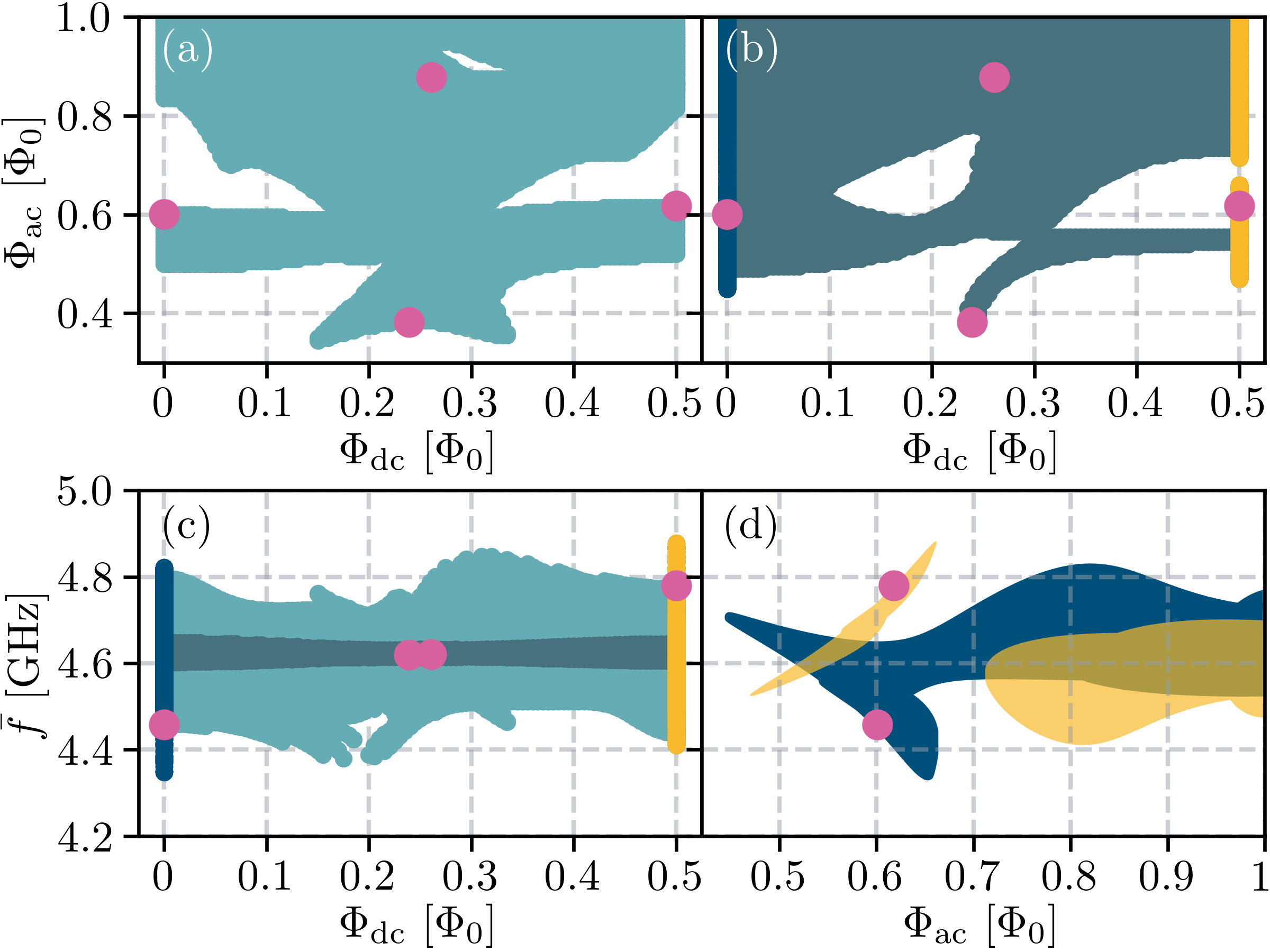}
	\caption{Top panels: Localization of the dynamical sweet spots in the $\Phi_\dc$--$\Phi_\ac$ plane for $p=2$ (a) and $p=3$ (b).
		We highlight the dynamical sweet spots when parking at the maximum (blue) and minimum (yellow) of the band in (b).
		Bottom panels: Corresponding time-averaged frequency as a function of the parking flux (c) and as a function of the modulation amplitude (d) with the same color code.
		The pink dots correspond to the results for a monochromatic modulation, $\alpha=0$.
		Same transmon parameters as in Fig.~\ref{fig_findsweetspot}.}
	\label{fig_sweetspots}
\end{figure}

Parking the qubits at an extremum of the frequency band, at $\Phi_\dc=0,\ \frac{1}{2}\Phi_0$, is of particular interest.
First because it is a sweet spot without modulation and
second because the symmetry of the band around the parking flux provides additional properties.
When an odd harmonic $p$ is used the qubit is first order insensitive to additive slow flux noise, $\partial\bar{f}/\partial\Phi_\dc=0$, and the odd sidebands vanish, $F_{2k+1}^{(\ac)}=0$.
These two properties are reminiscent of the sweet spots under monochromatic modulation~\cite{Nico:2019}.
On the other hand, when an even harmonic $p$ is used, the qubit is first order insensitive to additive slow flux noise for $\theta=\pi/2$ and the time-averaged frequency is unchanged under $\theta\to\pi-\theta$.

The location of dynamical sweet spots and the time-averaged frequency at these points of operation are plotted in Fig.~\ref{fig_sweetspots}.
Starting from four sweet spots with monochromatic modulation (for a maximum modulation amplitude of $\Phi_\ac=\Phi_0$), the bichromatic modulation unveils a continuum of dynamical sweet spots.
The accessible time-averaged frequencies at these sweet spots span $60\,\%$ of the tunability range for $p=2$ and $65\,\%$ for $p=3$.
While most of this range is available at all parking flux for $p=2$, most of the flexibility is obtained at the maximum and the minimum of the band for $p=3$ ($60\,\%$ at each point). 
A convenient configuration is to park the qubit at an extremum of the frequency band where it is protected without modulation and to modulate with $p=3$ to activate entangling gates on a wide range of time-averaged frequencies while suppressing odd sidebands.

\begin{figure}[t]
	\includegraphics[width=\columnwidth]{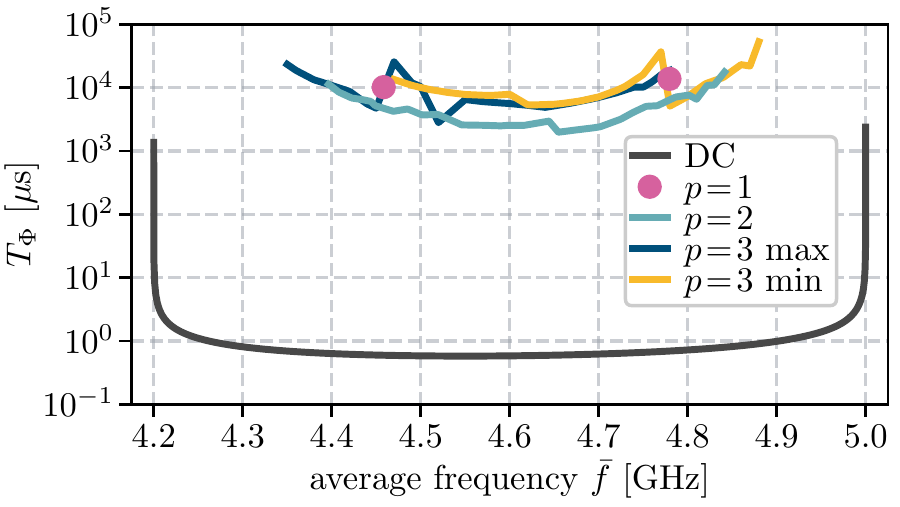}
	\caption{Dephasing time as a function of the time-averaged frequency for a constant DC bias (full line) and at dynamical sweet spots: monochromatic modulation ($p=1$, dots), bichromatic modulation with $p=2$ (cyan), $p=3$ around a maximum (blue) and around a minimum (yellow).
	The dephasing time is obtained by numerically averaging the coherent dynamics over realizations of $1/f$ flux noise for $A_\dc=A_\ac =33\,\mu\Phi_0$ at a modulation frequency $f_m=150\unit{MHz}$.
	The dynamical sweet spots are chosen to maximize the central sideband weight.
	Same transmon parameters as in Fig.~\ref{fig_findsweetspot}.}
	\label{fig_Tphi}
\end{figure}

The dephasing times of a tunable transmon in presence of strong $1/f$ flux noise is plotted in Fig.~\ref{fig_Tphi} for different flux biases.
The dephasing rate is calculated by averaging the coherent dynamics over $N_\mathrm{shot}=4000$ realizations of strong $1/f$ flux noise characterized by the strengths $A_\dc=A_\ac=33\,\mu\Phi_0$ on $1\unit{ns}$ time steps.
Under $1/f$ flux noise, the off-diagonal elements of the density matrix have a Gaussian decay, $e^{-(\Gamma_\phi t)^\beta}$ with $\beta\approx1.9$.
The exponent $\beta$ is defined by the $1/f$ behavior of the spectral density at low frequencies, above the infra-red cutoff $1/T_\mathrm{ir}$.
At dynamical sweet spots the dephasing time is limited by second order derivatives and by the noise spectrum around non-zero harmonics of the modulation frequency.
The low-frequency flux fluctuations are  not contributing to dephasing, as a result the off-diagonal density-matrix elements decay exponentially~\cite{Nico:2019}.

As can be seen in Fig.~\ref{fig_Tphi}, with a DC bias the dephasing time decreases by more than three orders of magnitude when the qubit is not parked at an extremum of the tunability band.
Dephasing times $T_\phi>2\unit{ms}$ are reached at sweet spots,
as large as $T_\phi\approx10\unit{ms}$ at the two dynamical sweet spots under a monochromatic modulation 
and between $T_\phi=2\unit{ms}$ and $T_\phi=50\unit{ms}$ over two thirds of the band using a bichromatic modulation.
Experimentally, the coherence time is limited by other sources of dephasing not taken into account in this study focused on protection from $1/f$ flux~noise.

\begin{figure}[t]
	\includegraphics[width=\columnwidth]{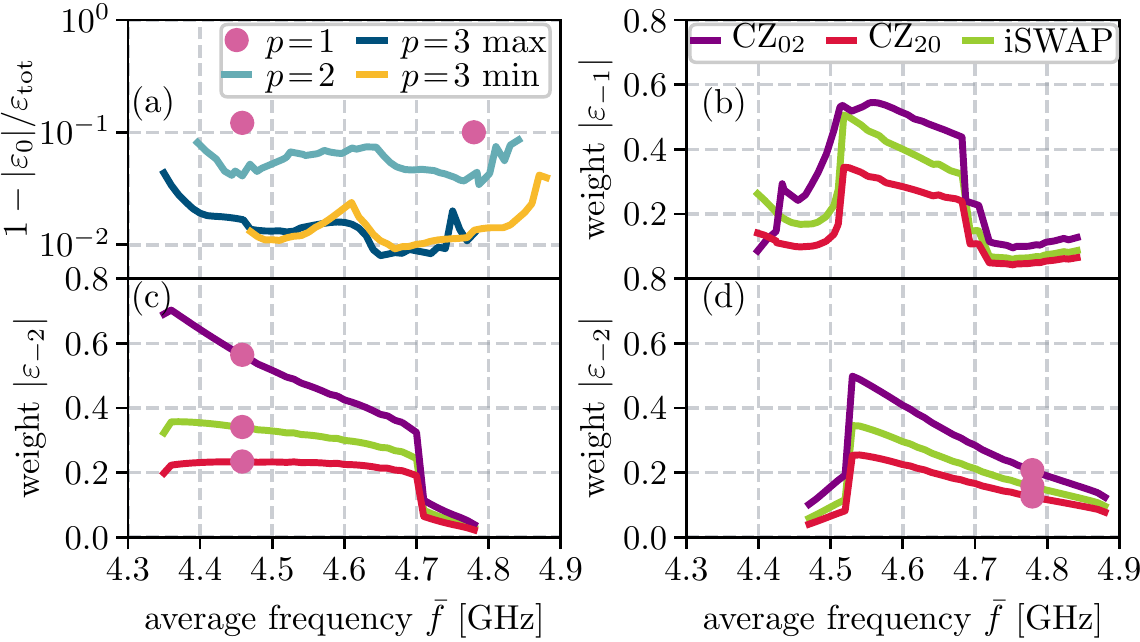}
	\caption{Maximum sideband weights as a function of the time-averaged frequency at dynamical sweet spots for $\Phi_\ac\leq0.75\,\Phi_0$.
		(a) Central sideband, $k=0$, for $f_m=250\unit{MHz}$.
		(b) Sideband $k=-1$ for $p=2$ and $0.2\,\Phi_0\leq\Phi_\dc\leq0.3\,\Phi_0$.
		The modulation frequency is chosen to activate parametric entangling gates between the tunable transmon of Fig.~\ref{fig_findsweetspot} and a fixed-frequency transmon characterized by $f_{F_{01}}=4.2\unit{GHz}$ and $f_{F_{12}}=4\unit{GHz}$.
		Bottom panels: Sideband $k=-2$ for $p=3$ at $\Phi_\dc=0$ (c) and $\Phi_\dc=0.5\,\Phi_0$ (d).
		Modulation frequency for parametric entangling gates with a fixed-frequency transmon characterized by $f_{F_{01}}=4\unit{GHz}$ and $f_{F_{12}}=3.8\unit{GHz}$.
	The dots correspond to the case of a monochromatic modulation.}
	\label{fig_weight}
\end{figure}

\section{Entangling gates protected from flux noise}
\label{Sec_gates}

Fast entangling gates can be activated by bringing two qubits or qutrits into resonance to allow coherent exchange between two states to take place during the desired time.
The capacitive coupling between superconducting qubits generates a coupling between states of the same parity.
When the coupling is strong but stays smaller than the transition frequencies, the rotating wave approximation is used to keep the interaction terms occurring between states of the same number of excitations.
From the capacitive coupling $g$ between a tunable transmon and a fixed-frequency transmon, one obtains a coupling of strength $g\mu_{01}$ between the states $\ket{01}$ and $\ket{10}$, a coupling $\sqrt{2}g\mu_{12}$ between $\ket{11}$ and $\ket{02}$, and a coupling $\sqrt{2}g\mu_{01}$ between $\ket{11}$ and $\ket{20}$ (notation $\ket{\mathrm{fixed},\mathrm{tunable}}$).
The terms $\mu_{01},\mu_{12}\approx1$ are flux dependent, they come from the nonlinearity of the Josephson junction when diagonalizing and are given in Appendix~\ref{App_mu} from perturbation theory~\cite{Nico:2018}.

\begin{figure}[t]
	\includegraphics[width=\columnwidth]{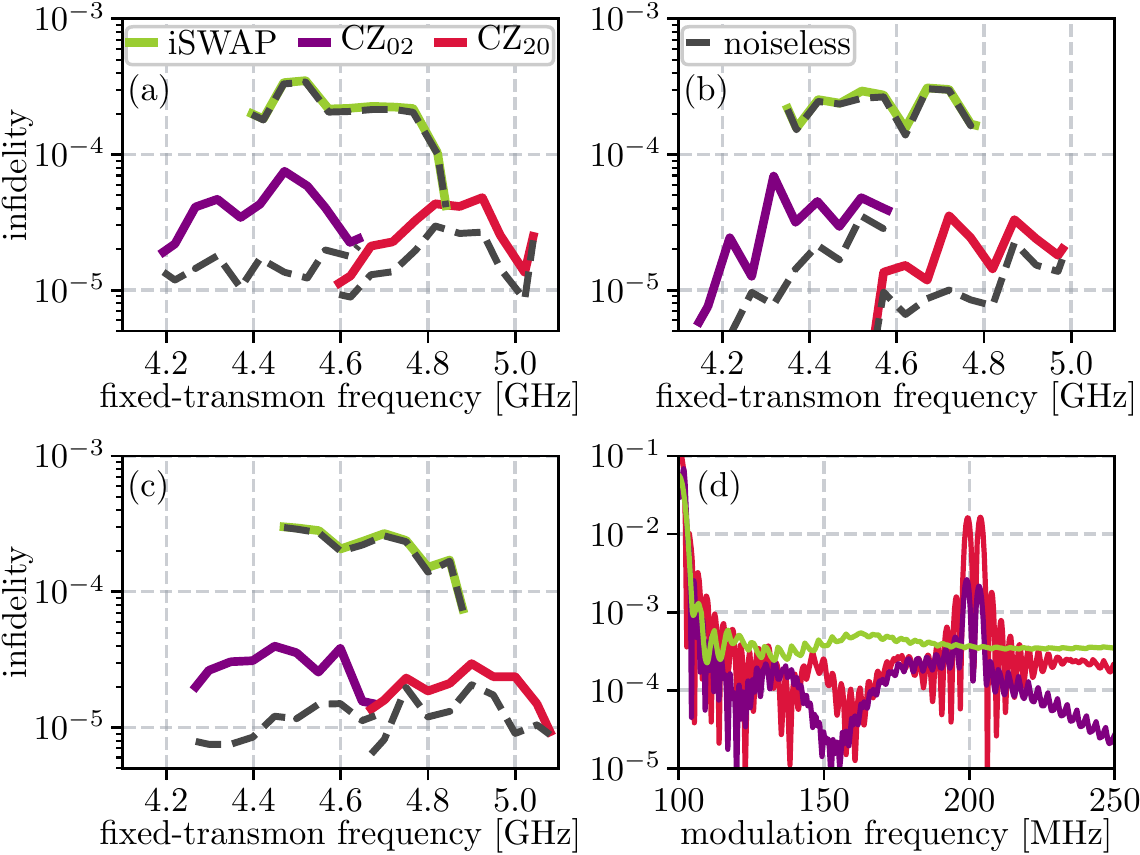}
	\caption{
		Infidelity of entangling gates operated at dynamical sweet spots.
		The time-averaged transition frequency of a tunable transmon under bichromatic modulation is brought on resonance with the transition frequency of a fixed-frequency transmon.
		The infidelity is plotted as a function of the fixed-transmon frequency for $p=2$ (a), $p=3$ at a maximum of the band (b) and $p=3$ at a minimum (c).
		The gate parameters (sideband weight, modulation frequency, gate time, local Z rotations) are optimized without flux noise (dashed lines).
		The dynamics is then averaged over realizations of $1/f$ flux noise (solid lines).
		Tunable transmon of Fig.~\ref{fig_findsweetspot} and fixed-frequency transmon characterized by $f_{F_{01}}=4\unit{GHz}$ and $f_{F_{12}}=3.8\unit{GHz}$
		with a coupling $g=2.5\unit{MHz}$.
		Same noise strengths as in Fig.~\ref{fig_Tphi}.
		High-fidelity entangling gates are accessible under $1/f$ flux noise for a fixed transmon frequency within the full tunability range.
		(d) Dependence of the infidelity against modulation frequency, without flux noise for $p=3$ at a maximum and a fixed-transmon frequency $f_{F_{01}}=4.57\unit{MHz}$.
	}
	\label{fig_infidelity}
\end{figure}

Bringing two qubits into resonance is readily performed with a DC flux pulse.
The tunable qubit is usually not protected from flux noise during the interaction time and high-fidelity gates are obtained with ultrashort gate times~\cite{Barends:2019,Arute:2019}.
To further improve the performance, flux modulation can be used as a protection against flux noise.
Under flux modulation, the charge operator is dressed with sidebands centered around the time-averaged frequency $\bar{f}$ and separated by the modulation frequency $f_m$~\cite{Nico:2018}.
The $k^\mathrm{th}$ sideband at $\bar{f}+kf_m$ is characterized by the weight $\varepsilon_k$.
The weights are defined from the time-dependence of the transition frequencies and the coupling terms $\mu_{01},\mu_{12}$.
When a tunable transmon is capacitively coupled to a fixed-frequency transmon, 
entangling gates can be realized by bringing the time-averaged transition frequency of the tunable transmon in resonance with the desired transition frequency of the fixed-frequency transmon.
The interaction in the one-excitation subspace is used to enact an iSWAP gate, obtained when $\bar{f}_{01}=f_{F,01}$ during the time $\tau_\mathrm{iSWAP}=1/(4|\varepsilon_0|g)$.
The interaction in the two-excitation subspace temporarily involves states out of the computational basis to realize a Controlled-Z (CZ) gate, obtained at $\bar{f}_{10}=f_{F,12}$ ($\CZ_{02}$) or $\bar{f}_{12}=f_{F,01}$ ($\CZ_{20}$) during the time $\tau_{\CZ}=1/(2\sqrt{2}|\varepsilon_0|g)$.

The gate is finally optimized by choosing the dynamical sweet spots that maximize the weight of the central sideband, achieved when $|\varepsilon_0|$ reaches the maximal weight value $\varepsilon_\mathrm{tot}$, see Appendix~\ref{App_weight}.
The central sideband weight at the sweet spots is plotted as a function of the time-averaged frequency in Fig.~\ref{fig_weight}~(a) for a modulation frequency $f_m=250\unit{MHz}$,
where weights $|\varepsilon_0| > 0.99\,\varepsilon_\mathrm{tot}$ are achievable.
For large enough modulation frequencies, that is when the pulse duration comprises several periods, the time-averaged frequency does not depend on the modulation frequency.
The weight of the central sideband however depends on~$f_m$, it converges to the maximum value
$\varepsilon_\mathrm{tot}$
at large modulation frequencies.

The infidelity of entangling gates is plotted in Fig.~\ref{fig_infidelity} for $p=2$ and $p=3$ as a function of the fixed transmon frequency $f_{F,01}$ for a coupling $g=2.5\unit{MHz}$.
In the numerical simulations, the modulation frequency, the gate time and the local $Z$ rotations are optimized.
A typical dependence of the infidelity on the modulation frequency is shown in Fig.~\ref{fig_infidelity}~(d).
Infidelities lower than $5\times10^{-4}$ are achieved on a wide range of modulation frequencies, away from resonances with other parametric gates.
For instance when $\CZ_{02}$ is activated with the central sideband, $\bar{f}_{12}=f_{F_{01}}$, then iSWAP is activated at the modulation frequency $f_m=|\bar{f}_{01}-f_{F_{01}}|/k=\bar{\eta}/k$ and $\CZ_{20}$ at $f_m=|\bar{f}_{01}-f_{F_{12}}|/k=(\bar{\eta}+\eta_F)/k$.
In the case of iSWAP, $\CZ_{02}$ is activated at $f_m=\bar{\eta}/k$ and $\CZ_{20}$ at $f_m=\eta_F/k$.
These resonance conditions appears in Fig.~\ref{fig_infidelity}~(d) around $f_m=100\unit{MHz}$ and $f_m=200\unit{MHz}$ for CZ gates and $f_m=100\unit{MHz}$ for iSWAP since odd sideband weights vanish for $p=3$ around a maximum.

The optimal parameters are used to calculate the fidelity of the entangling gates in the presence of strong $1/f$ flux noise by numerically averaging the coherent dynamics over $N_\mathrm{shot}=4000$ realizations of $1/f$ flux noise.
The average process fidelity, with respect to the ideal gate $\hat{V}$, of the evolution operator $\hat{U}_j(\tau)$ at gate time $\tau$ for shot number $j$ is equal to,
\begin{align}
F_\mathrm{avg}&=\frac{\sum_{j=1}^{N_\mathrm{shot}}F_j}{N_\mathrm{shot}}, &
F_j&=\frac{|\tr\{\hat{V}^\dag \hat{\Pi} \hat{U}_j(\tau)\hat{\Pi}\}|^2+d}{d^2+d},
\end{align}
with $d=4$ and $\hat{\Pi}$ the projector on the computational basis.
The noise strengths are the same as in Fig.~\ref{fig_Tphi}, $A_\dc=A_\ac =33\,\mu\Phi_0$.
As shown in Fig.~\ref{fig_infidelity}, it is possible to find high-fidelity two-qubit gates inside the whole tunability range of the qubit.
The CZ infidelity saturates around $7\times10^{-5}$, compatible with the coherence-limited infidelity for the dephasing times $T_\phi>2\unit{ms}$ of Fig.~\ref{fig_Tphi}.
The iSWAP infidelity is mainly limited by the coherent errors on the $\ket{11}$ state that accumulates a phase $\sim4^\circ$ due to the dispersive interactions in the two-excitation subspace.

Such entangling gates, protected from flux noise, can be implemented between tunable qubits.
They are realized by operating both tunable qubits at dynamical sweet spots, chosen to satisfy the resonance condition between the time-averaged transition frequencies.

The flexibility brought by the bichromatic modulation can also be used to optimize the performance of parametric entangling gates using satellite sidebands~\cite{Nico:2018,Shane:2018,Matt:2018,Sab:2019,Deanna:2019b}.
To activate the coherent exchange that generates the two-qubit gate with the $k^\mathrm{th}$ sideband ($k\neq0$) the flux bias is modulated at the frequency $f_m=|\bar{\Delta}/k|$.
Here, $\bar{\Delta}$ is the time-averaged detuning between the two states used in the coherent exchange.
Explicitly, 
$\bar{\Delta}=\bar{f}_{01}-f_{F_{01}}$ for iSWAP, 
$\bar{\Delta}=\bar{f}_{12}-f_{F_{01}}$ for $\CZ_{02}$,
$\bar{\Delta}=\bar{f}_{01}-f_{F_{12}}$ for $\CZ_{20}$.
The sideband weight depends on the modulation frequency, it typically oscillates at low modulation frequencies and then vanishes at high frequencies.
The gate time is inversely proportional to the sideband weight, and since the modulation frequency is set by the time-averaged detuning, it is possible to reduce the gate time by optimizing the dynamical sweet spot parameters.
The weight of the sideband $k=-1$ at dynamical sweet spots for $p=2$ is plotted as a function of the time-averaged frequency in Fig.~\ref{fig_weight}~(b)
and the weights for $k=-2$ and $p=3$ are plotted in the bottom panels of Fig.~\ref{fig_weight}.
The modulation frequency is chosen to activate an iSWAP or a CZ gate between the tunable transmon and a fixed-frequency transmon.
The weight $|\varepsilon_{-2}|$ can be optimized with respect to the case of a monochromatic modulation, in particular in this configuration when the tunable qubit is parked at the minimum of the band.

Furthermore, the variety of effective transmon parameters accessible at dynamical sweet spots can be used to move the modulation frequency away from collisions with other sidebands when necessary.
Dynamical sweet spots moreover provide a robustness against slow drifts of control parameters.
In addition to gate optimization, sideband engineering can be used for error mitigation using Richardson's extrapolation.
Indeed, as shown in Fig.~\ref{fig_weight}, the gate time can be varied by changing the sideband weight at dynamical sweet spots.
The effect of qubit decay and other sources of dephasing can then be circumvented by extrapolating down to zero noise the result of the algorithm of interest~\cite{Temme:2017, Kandala:2019}.

\section{Qubit manipulation at dynamical sweet spots}
\label{Sec_TLS}

When not used in entangling gates, the tunable qubits can be parked at a sweet spot with no modulation located at a maximum or a minimum of the frequency band.
It is however sometimes not possible due to the presence, for instance, of two-level systems (TLS) close to these extrema~\cite{Arute:2019}.
Under modulation when the central sideband weight tends to its maximum value, the other sideband weights vanish and the qubit behaves closely to an undriven qubit defined by the time-averaged transition frequency~\cite{Li:2013}.
At a dynamical sweet spot with a time-averaged frequency sufficiently away from the TLS frequency, the qubit is affected by neither slow flux noise nor by the TLS.

\begin{figure}[t]
	\centering
	\includegraphics[width=0.9\columnwidth]{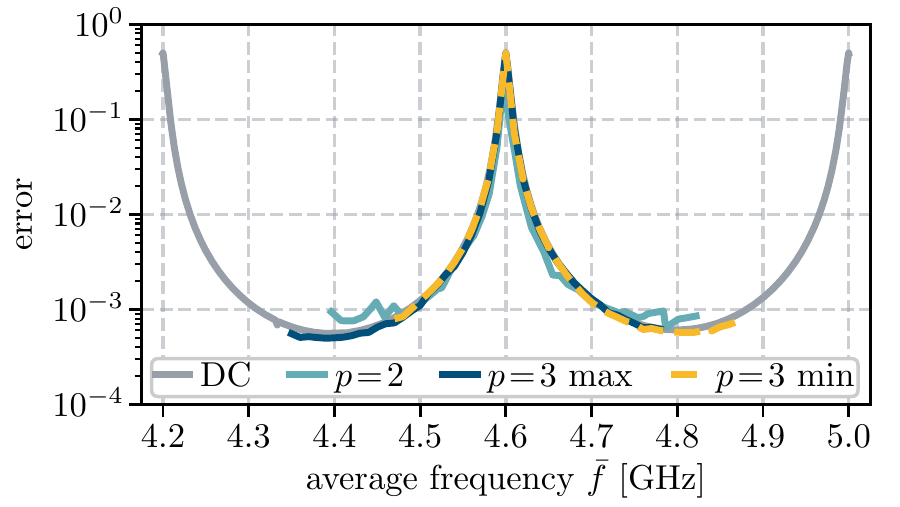}
	\caption{Maximum error with respect to identity for a tunable qubit coupled to three two-level systems: at the maximum frequency, at the minimum frequency and in the middle of the band with a coupling of $3\unit{MHz}$.
	The leakage to the TLS at dynamical sweet spots follows the error obtained with a DC-biased transmon.
	The case $p=2$ is slightly less efficient due to the lower values of the central sideband weights obtained in Fig.~\ref{fig_weight}~(a).}
	\label{fig_TLS}
\end{figure}

This is illustrated in Fig.~\ref{fig_TLS}.
The tunable qubit is coupled to three TLS, one at the maximum, one at the minimum and one in the center of the band, with a coupling strength of $3\unit{MHz}$ each.
We then compute the maximum infidelity with respect to identity at various operating points (over $1\,\mu\mathrm{s}$).
Even though the tunable qubit is periodically driven through the TLS anticrossings, the error for a bichromatic modulation follows the one of a DC bias.
This result illustrates the fact that a qubit under modulation behaves like a undriven qubit when the central sideband weight is maximized.

Similarly, it is possible to perform single qubit gates under modulation by driving the qubit at the time-averaged frequency.
Dispersive qubit measurement is possible as well under modulation, the dispersive shift can even be engineered by choosing a dynamical sweet spot with optimal time-averaged detuning with the readout resonator.
In both cases, the modulation frequency can be optimized if necessary, for instance to avoid that the time-averaged frequency or detuning with the resonator is close to a sideband frequency.
As a conclusion, tunable qubits can be operated at dynamical sweet spots to avoid both collisions and dephasing from slow flux noise.

\section{Conclusion}

We have shown how to combine the versatility of flux control of tunable superconducting qubits with the protection from slow flux noise of dynamical sweet spots.
The use of bichromatic flux modulation gives access to a continuum of dynamical sweet spots with a wide range of time-averaged frequencies and sideband weights.
Tunable qubits are operated at dynamical sweet spots optimized with sideband engineering to cope with imperfections, such as coupling to TLS defects, and to implement high-fidelity two-qubit gates, by substantially increasing the coherence-limited fidelities.
The diversity of the effective parameters at dynamical sweet spots under multi-harmonic flux modulation paves the way for optimal control of flux-activated entangling gates in superconducting quantum processors.

\begin{acknowledgments}
We thank Shoumik Chowdhury, Prasahnt Sivarajah and Deanna Abrams for the initial experimental exploration of dynamical sweet spots under bichromatic flux modulation.
We thank Deanna Abrams, Colm Ryan and Marcus Silva for useful discussions.
We thank Colm Ryan for critically reading the manuscript.
\end{acknowledgments}

\appendix

\section{Fourier series under bichromatic modulation}
\label{App_fourier}

Inserting the flux bias Eq.~\eqref{flux} into the tunable qubit frequency band Eq.~\eqref{FBA} yields the following series,
\begin{align}
f(t) &= \bar{f} + \sum_{k=1}^\infty \sum_{l\in\mathbb{Z}} \nu_{k,l} \cos(2\pi kf_mt+l\theta),\\
\nu_{k,l} &= 2 \sum_{n=0}^\infty \FDC_n \cos[n\phi_\dc+(p+1)l\tfrac{\pi}{2}-k\tfrac{\pi}{2}] \nonumber\\ 
&\hspace{16mm} \times \BJ_{pl-k}(n\phi_\ac\cos\alpha) \BJ_{l}(n\phi_\ac\sin\alpha),
\end{align}
with $\bar{f}$ of Eq.~\eqref{favg}.
Eq.~\eqref{fourier} is then obtained from,
\begin{align}
\FAC_k &= |z_k|, &
\theta_k &= \arg z_k, &
z_k &= \sum_{l\in\mathbb{Z}} \nu_{k,l} e^{il\theta}.
\end{align}

\section{Coupling under modulation}
\label{App_mu}

The flux dependence of the coupling strength in the transmon eigenbasis can be obtained in the transmon regime from perturbation theory in the small parameter $\xi(\Phi)=\sqrt{2E_C/E_J(\Phi)}$ with the charging energy $E_C$ and the effective Josephson energy 
$E_J(\Phi)=\sqrt{E_{J_1}^2+E_{J_2}^2+2E_{J_1}E_{J_2}\cos(2\pi\Phi/\Phi_0)}$ defined from the Josephson energies $E_{J_1}$ and $E_{J_2}$ of the SQUID loop.
If the coupling strength $g$ is defined at the flux $\Phi^*$, like at a maximum $\Phi^*=0$ in this study, the expansion of $\mu_{01}$ and $\mu_{12}$ in $\xi(\Phi)$ is equal to
\begin{align}
\mu_{01}(\xi)&=\zeta_{01}(\xi)/\zeta_{01}(\xi^*),&
\mu_{12}(\xi)&=\zeta_{12}(\xi)/\zeta_{12}(\xi^*),
\end{align}
where $\xi^*=\xi(\Phi^*)$, such that $\mu(\xi^*)=1$.
The parameters $\zeta$ are equal to~\cite{Nico:2018},
\begin{align}
\zeta_{01}\approx \frac{1}{\sqrt{\xi}}\bigg[&
1
-\frac{1}{2^{3}}\xi
-\frac{11}{2^{8}}\xi^2
-\frac{65}{2^{11}}\xi^3
-\frac{4203}{2^{17}}\xi^4\nonumber\\
&-\frac{40721}{2^{20}}\xi^5
-\frac{1784885}{2^{25}}\xi^6
+\dots\bigg],\\
\zeta_{12}\approx \frac{1}{\sqrt{\xi}}\bigg[&
1
-\frac{1}{2^{2}}\xi
-\frac{73}{2^{9}}\xi^2
-\frac{79}{2^{9}}\xi^3
-\frac{113685}{2^{19}}\xi^4\nonumber\\
&-\frac{747533}{2^{21}}\xi^5
-\frac{175422349}{2^{28}}\xi^6
+\dots\bigg].
\end{align}

\section{Sideband weights}
\label{App_weight}

The sideband weights are obtained from the time dependence of the charge operator in the interaction picture,
which is Fourier expanded as follows,
\begin{align}
\mu(t)e^{i\int_0^t\mathrm{d}t'f(t')} = \sum_{k\in\mathbb{Z}}\varepsilon_ke^{i(\bar{f}+kf_m)t}.
\end{align}
The maximal value of a sideband weight, $\varepsilon_\mathrm{tot}$, is obtained when all the others vanish.
Using Parseval's theorem we get
\begin{align}
\varepsilon_\mathrm{tot}=\sqrt{\int_0^1\mathrm{d}x\,\mu^2(x/f_m)}.
\label{maxweight}
\end{align}


\begin{thebibliography}{64}
\bibitem [{Fou()}]{FourierDC}%
	\BibitemOpen
	\href@noop {} {}\bibinfo {note} {Equivalently, the Fourier coefficients are
		equal to $F^{(\protect \mathrm {dc})}_{0,1,2,3,4,5,6\protect \dots } =
		4.62\protect \tmspace +\thinmuskip {.1667em}\protect \mathrm {GHz}$,
		$398\protect \tmspace +\thinmuskip {.1667em}\protect \mathrm {MHz}$,
		$-24.7\protect \tmspace +\thinmuskip {.1667em}\protect \mathrm {MHz}$,
		$2.38\protect \tmspace +\thinmuskip {.1667em}\protect \mathrm {MHz}$,
		$-271\protect \tmspace +\thinmuskip {.1667em}\protect \mathrm {kHz}$,
		$33.7\protect \tmspace +\thinmuskip {.1667em}\protect \mathrm {kHz}$,
		$-4.42\protect \tmspace +\thinmuskip {.1667em}\protect \mathrm {kHz}\protect
		\dots $}\BibitemShut {NoStop}%
\end{thebibliography}
\end{document}